\documentclass[useAMS,usenatbib,usegraphicx]{mn2e}
\usepackage[utf8]{inputenc}
\usepackage{times}
\usepackage{microtype}
\usepackage{paralist}
\usepackage{amsmath}
\usepackage{natbib}
\usepackage{graphicx}
\usepackage{epsfig}
\usepackage{txfonts}
\usepackage{color}
\usepackage{multirow}
\usepackage{morefloats}
\usepackage{amssymb}
\usepackage{epsfig}
\usepackage{xspace}

\newcommand{\gro}{GRO~J1744$-$28\xspace}

\title[{\emph{BeppoSAX} observations of \gro}]{\emph{BeppoSAX} observations of \gro} 
\author[R.\,Doroshenko et. al]{R.\,Doroshenko$^1$, A.\,Santangelo$^1$, V.\,Doroshenko$^1$, V.\,Suleimanov$^{1,2}$, S.\,Piraino$^{1,3}$\\
$^1$Institut für Astronomie und Astrophysik, Kepler Center for Astro and Particle Physics, Sand 1, 72076 Tübingen, Germany\\
$^2$Kazan (Volga region) Federal University, Kremlevskaya str. 18, 42008 Kazan, Russia\\
$^3$INAF – IASF di Palermo, via Ugo La Malfa 153, 90146 Palermo, Italy}

\begin{document} 
\maketitle
\begin{abstract}
We present an analysis of \emph{BeppoSAX} observations of the unique transient
bursting X-ray pulsar \gro. The observations took place in March 1997 during the decay
phase of the outburst. We find that the persistent broadband X-ray continuum of
the source is consistent with a cutoff power law typical for the accreting
pulsars. We also detect the fluorescence iron line at $6.7$\,keV and an
absorption feature at $\sim$\,4.5\,keV, which we interpret as a cyclotron line.
The corresponding magnetic field strength in the line forming region is $\sim
3.7\times 10^{11}$\,G. Neither line is detected in the spectra of the bursts.
However, additional soft thermal component with $kT \sim 2$\,keV was required
to describe the burst spectrum. We briefly discuss the nature of this component and
argue that among other possibilities it might be connected with thermonuclear
flashes at the neutron star surface which accompany the accretion-powered
bursts in the source.
\end{abstract}

\section{Introduction} The peculiar transient X-ray source \gro also known as
the bursting pulsar, was discovered on December~2,~1995, with the Burst and
Transient Source Experiment (\emph{BATSE}), on board the Compton Gamma Ray
Observatory (\emph{CGRO}) \citep{Fishman:1995p6622, Kouveliotou:1996p6636}.
Three major outbursts have been observed since the discovery. The first lasted
from the end of 1995 until the beginning of 1996, the second from December 1996
until April 1997, and the third from January to May 2014
\citep{Degenaar.etal:14, Younes.etal:15, Dai.etal:15}. The peak outburst
luminosity in X-rays reaches $L_{\rm X}\sim 10^{37} - 10^{38}$~erg~s$^{-1}$
\citep{Woods:1999p6285}, while the quiescent luminosity is $\sim L_{\rm X} \sim
10^{33}$~erg~s$^{-1}$ \citep{Wijnands:2002p6217, Daigne:2002p6221,
Degenaar:2012p7088}. The source exhibits coherent pulsations with a period of
$\sim 0.467$~s, associated with the rotation of the neutron star i.e. is an
accreting X-ray pulsar in a binary system. \cite{Finger:1996p6585} determined
the orbital period and semimajor axis to be $\rm P_{\rm orb} = 11.8$~days and $
a = 1.12$~R$_{\odot}$ ($\sim 7.8 \times 10^{10}$~cm) respectively. The mass
function was estimated to be $\rm f(M) = 1.36 \times 10^{-4}$~M$_{\odot}$
implying, for the canonical neutron star mass of $\rm M_{\rm
NS}$~=~1.4~M$_{\odot}$, accretion via Roche lobe overflow from a strongly
evolved red giant remnant with mass about $0.2 - 0.7$~M$_{\odot}$
\citep{Daumerie:1996p6674, Miller:1996p6366, Sturner:1996p6393, BB:97,
Rappaport:1997p6322}. The photoelectric absorption does not change significantly during
the outbursts and is comparable with interstellar absorbtion $\rm N_{\rm H}\sim
(5 - 6) \times 10^{22}$~cm$^{-2}$ in the direction of the Galactic center where
the source is thought to be located at distance of $\sim$8~kpc
\citep{Dotani:1996p6654, Nishiuchi:1999p6279}.

Probably the most unusual property of \gro is the relatively short $\sim10$\,s
bursts observed from the source. Since the discovery, several thousands of bursts have been detected, all at high luminosity \citep{Nishiuchi:1999p6279}. During the
bursts the flux increases by an order of magnitude, and it is often followed by a drop to below
the pre-burst level for several tens of seconds to minutes depending on the
burst fluence \citep{Lewin:1996p469,Nishiuchi:1999p6279, Younes.etal:15}.
Pulsations are observed during the bursts, although with a phase
shift of $\sim 5$\% with respect to the persistent emission
\citep{Kouveliotou:1996p6636,Strickman:1996p6385}.

A similar bursting activity has been observed only in
another source the transient LMXB MXB~1730$-$335, usually referred to as the
Rapid Burster (RB) \citep{Lewin:1976p6792, LvPT:93}. For both sources, it has
been suggested that the origin of the bursts could be due to accretion flow
instabilities which intermittently enchance the accretion rate onto the neutron
star (so-called Type II bursts). However details are still unclear. For
instance, unlike the RB, Type I bursts, associated with thermonuclear flashes
on the surface of neutron star, are generally thought not to occur in \gro. On
the other hand, contrary to the RB case, the duration of the \gro bursts is
remarkably stable ($\sim10$\,s), and a large fraction of the bursts exhibit a
temporal profile characterized by a fast rise followed by an exponential decay
typical of classical bursters. Type II bursts in RB have rather
irregular profiles. These differences are likely related to the different magnetic
field strengths of the neutron star companion, which is expected, given the presence of
pulsations, to be stronger in \gro. In the case of \gro,
\cite{Finger:1996p6585} reported an upper limit on the magnetic field strength of $B\le 6 \times 10^{11}$~G. This estimate is based on the requirement that the plasma at the inner
edge of the accretion disk (disrupted by the magnetosphere) moves faster than
the magnetic field lines as otherwise the accretion would be centrifugally
inhibited (so-called ``propeller'' regime, \citealt{illarionov75}). Later
\cite{Cui:1997p6334} found a possible transition to the ``propeller'' regime in
\emph{RXTE} data at lower luminosity, which allowed for an estimate of the the magnetic field at $\sim 2.4 \times 10^{11}$~G. Recently \citet{Dai.etal:15} reported the
observation of an absorption feature at E$\sim$\,4.7\,keV, during the 2014
outburst of the source. Interpreting the line as due to cyclotron resonance
scattering, the authors estimated a magnetic field of $\sim 5.3\times
10^{11}$\,G in good agreement with earlier estimates.

In this paper we report on independent discovery of the same feature in
\emph{BeppoSAX} observations carried out during the outburst in 1997
(Sect.\,3.2)\footnote{We learned of the content by D'ai et al when the manuscript was in the final steps of preparation.}
We analyze three \emph{BeppoSAX} observations of \gro carried out
during the outburst in 1996-1997 and present a detailed spectral and timing
analysis of the persistent flux and the bursts. We also briefly discuss the origin
of the bursts in \gro. The instruments and methods of the analysis are
described in Section\,2. Results from the timing and spectral analysis are
summarised in Section\,3. In Section\,4, we discuss the observational results
and finally provide a summary of the paper in Section\,5.

\section{Observations}

\emph{BeppoSAX} was an Italian X-ray astronomy satellite, with Dutch participation, which
operated from 1996 to 2002 \citep{Boella:1997p3749}. It covered a broad energy
band 0.1--300\,keV with four Narrow Field Instruments (NFIs) and two Wide
Field Cameras \citep{Jager:1997p6454}. The NFIs included the Low-Energy
Concentrator Spectrometer \citep[LECS, 0.1--10 keV,] []{Parmar:1997p3751},
three identical Medium-Energy Concentrator Spectrometers \citep[MECS,
2--10\,keV,][] {Boella:1997p3748}, and two collimated high energy detectors
with good energy resolution and low instrumental background: the High Pressure
Gas Scintillation Proportional Counter \citep[HPGSPC, 4--120\,keV, FWHM energy
resolution of 8\% at 10 keV and 5.5\% at 20 keV,][]{Manzo:1997p3754} and the
Phoswich Detection System \citep[PDS, 15--300\,keV, FWHM energy resolution of
24\% at 20 keV, and 14\% at 60\,keV,][]{Frontera:1997p3756}.

\begin{figure}
	\includegraphics[width=0.5\textwidth]{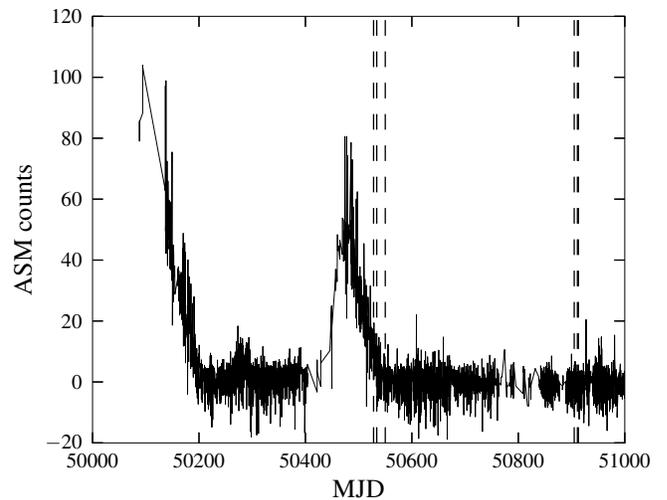}
	\caption{The light curve of GRO~J1744-28 observed by 
	the ASM onboard \emph{RXTE}. Dashed lines mark the \emph{BeppoSAX} 
	observations (Table\,\ref{tab:data}).}
	\label{fig:asm_lc}
\end{figure}

 \gro was observed by \emph{BeppoSAX} several times from March 1997 to April
1998 (Table~\ref{tab:data}). In
Figure~\ref{fig:asm_lc} the long term source light curve observed by the All
Sky Monitor onboard the Rossi X-ray Timing Explorer \citep[ASM
\emph{RXTE},][]{Bradt:1993p7} with superimposed \emph{BeppoSAX} observations is
presented. The first three observations of \emph{BeppoSAX} were performed in
the declining edge of the 1997 outburst, with a total exposure time for MECS of
approximately 270\,ks. In our work we use LECS, MECS, and PDS in 0.7 -- 4 keV,
2 -- 10 keV and 15 -- 120 keV energy ranges respectively. The HPGSPC was,
unfortunately, not operating during the first three observations. In the last
three observations, the source was in quiescence and below the sensitivity
thresholds of \emph{BeppoSAX}, so it was not detected. The data from the
quiescence, however, proved very useful for the background subtraction of the
outburst PDS data.

The standard \emph{BeppoSAX} pipeline was used for the processing of the data.
HPGSPC and PDS were operated in rocking mode to monitor local background and
its variation along the orbit. The source counts for LECS and MECS were
extracted from source centered circles with radii of 4 arcmin, while the
background from an annulus with outer radius of 18 arcmin. A detailed description
of \emph{BeppoSAX} data reduction procedures can be found in the handbook for
\emph{BeppoSAX}.

\begin{table}
	\caption{Observations of the X-ray source GRO~J1744-28 by 
	\emph{BeppoSAX}. Luminosity is estimated using unabsorbed flux  in the energy range 2--10 keV (MECS)  for  the distance $D = 8$ kpc.}
\begin{center}
\begin{tabular}
	{ccccc}
\hline\hline
MJD   &Exposure&   Period, s & $L_x, 10^{37}$ erg s$^{-1}$ & Orbital\\
	  &time, ks&             & 2--10 keV                   &phase$^a$\\
	\hline
              
50528 & 117.5  & 0.467044(1) & 2.8  & 0.9\\
50534 & 101.9  & 0.467044(1) & 1.8  & 0.4\\
50550 & 51.6   & 0.46705(1)  & 0.46 & 0.8\\
50905 & 62.5   &\multicolumn{2}{c}{source in}& 0.8\\
50911 & 34.5   &\multicolumn{2}{c}{quiescent state}& 0.3\\
50913 & 67.5   &             &      & 0.5\\
	\hline
\end{tabular}
$^a${Ephemeris from \cite{Kouveliotou:1996p6636}}
\end{center}	
\label{tab:data}
\end{table}

\section{Data analysis and results.} 
\subsection{Timing analysis.} 
\begin{figure}
	\includegraphics[width=0.45\textwidth]{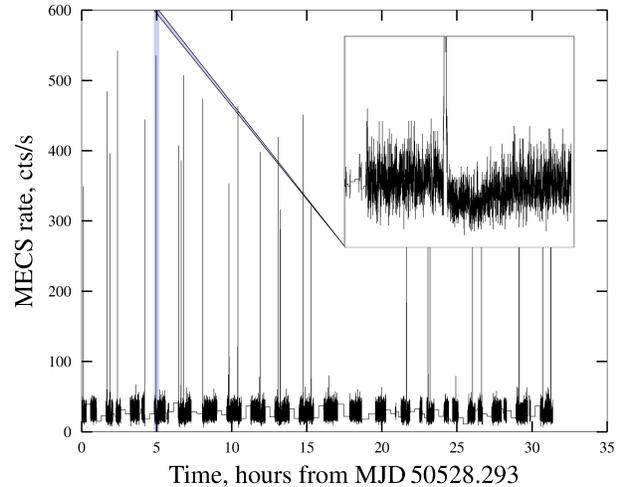}
	\caption{MECS light curve with bursts of the pulsar 
	\gro in 1997-03-21 with 2 s bin time in 2 -- 10 keV energy range. 
	On the zoom picture the flux drop after the burst is shown.}
	\label{fig:97lc}
\end{figure}

\begin{figure*}
	\includegraphics[width=\textwidth]{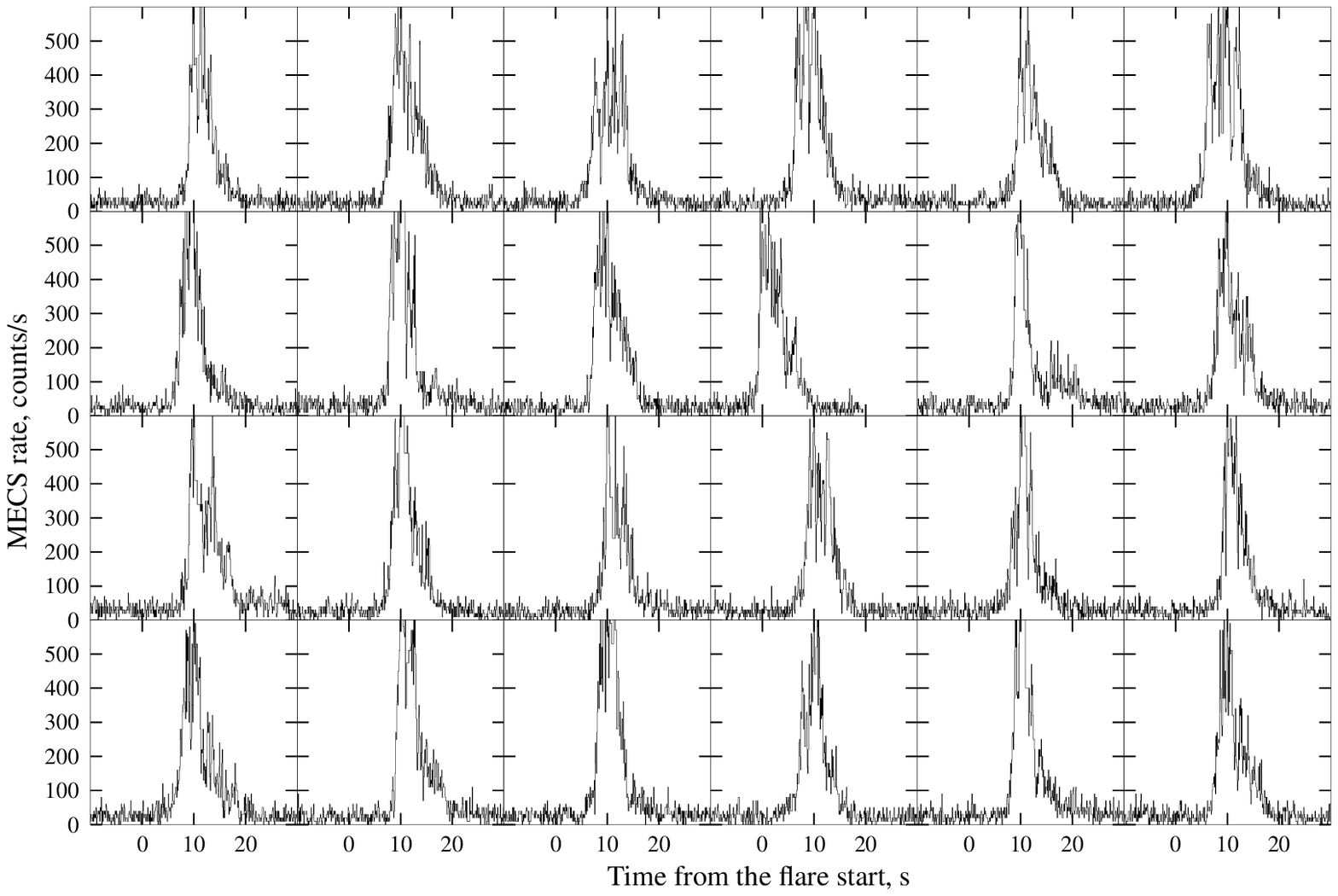}
	\includegraphics[width=\textwidth]{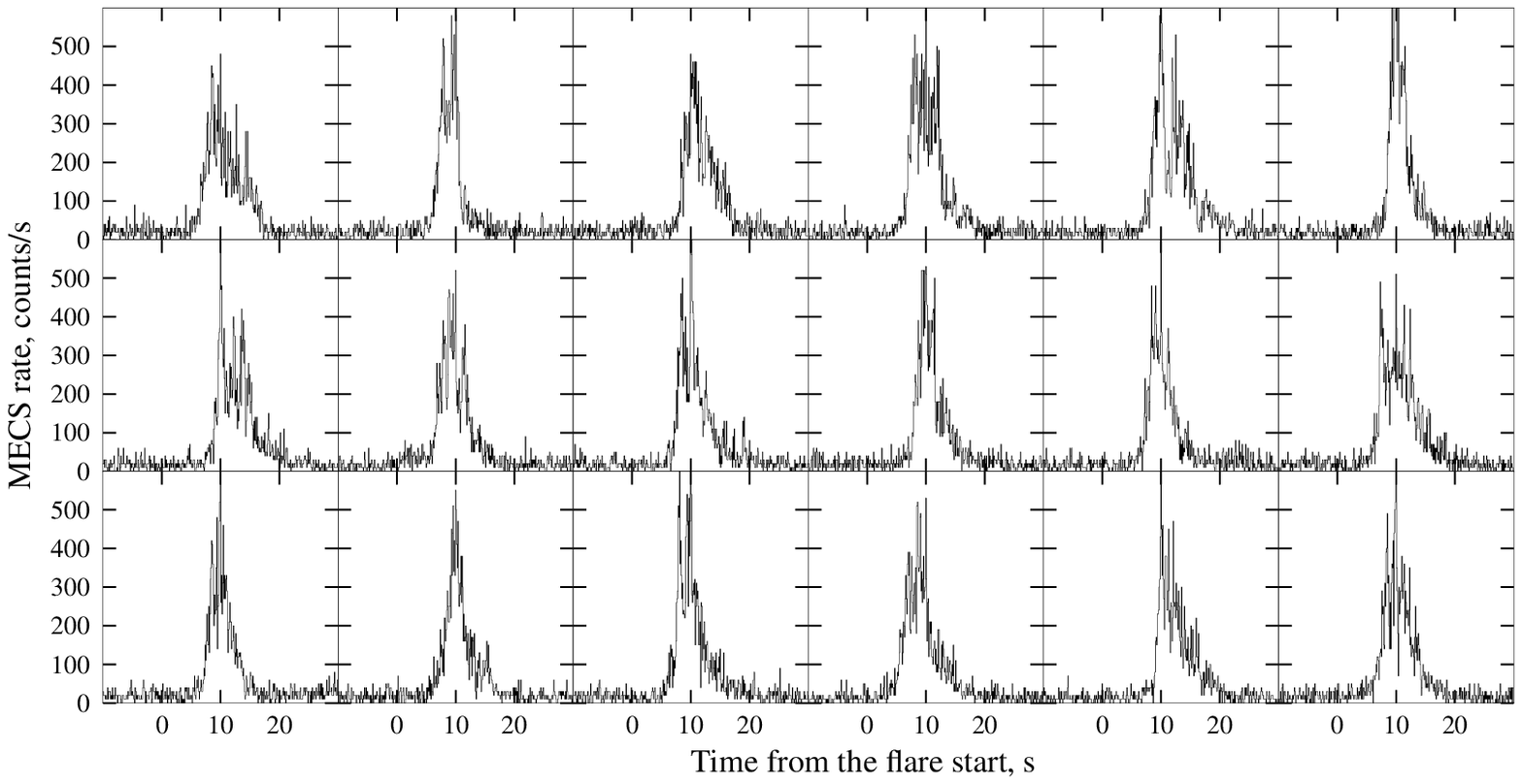}
	\caption{Light curves of all burst of the pulsar \gro in 1997-03-21 (top)       and 1997-03-27 (bottom), MECS.}
	\label{fig:bursts}
\end{figure*}

The 2--10 keV MECS light curve of the first observation (MJD 50528) is
presented in Fig.\ref{fig:97lc}. Bursts are clearly visible in the first two
observations, whereas none is detected in the last observation carried out
close to the end of the outburst at significantly lower luminosity. We detected
25 and 17 major bursts in the first two observations respectively, with a flux
increase of a factor of 10 or more. Typical burst duration and the interval between
the bursts are $\sim$10\,s and $\sim$ 2000\,s respectively (see also
Fig.~\ref{fig:bursts}, where all detected bursts are reported). The inter-burst
to burst fluence ratio in the energy range 2-20 keV is between 3 and 18, i.e.
compatible with earlier reports \citep{Lewin:1996p469,Jahoda97}. The burst profiles
exhibit variety of shapes: most are rather symmetric and in this respect are
similar to Type~II bursts observed in RB \citep{Lewin:1976p6792}. However,
about twenty percent (for example, the next to last one in
Fig.~\ref{fig:bursts}) exhibit the characteristic exponential decay typical of
thermonuclear Type~I bursts. The source also exhibited a few ``smaller''
bursts, a factor of $\sim 6$ less luminous than the major bursts. These have
already been reported by \cite{Nishiuchi:1999p6279} based on the \emph{ASCA}
data. We confirm the presence of the low luminosity bursts, although the
statistical quality of the data is not sufficient for any detailed analysis.

As reported earlier by several authors \citep{Lewin:1996p469, Younes.etal:15},
a flux drop following the burst is a relative common feature in \gro. Such a drop is observed in more than half of
the bursts in our sample. In Fig.~\ref{fig:97lc} a fragment of the light
curve in the 2--10\,keV is presented. The decrease of the mean flux from the
source after the burst, for about 100 sec, is evident. Note that a similar flux
depression is observed after Type II bursts of the RB and is usually
associated with the depletion of the inner edge of the accretion disk after
the bursting event due to enhanced accretion \citep{Lewin:1996p469}.

\begin{figure}
	\includegraphics[width=0.5\textwidth]{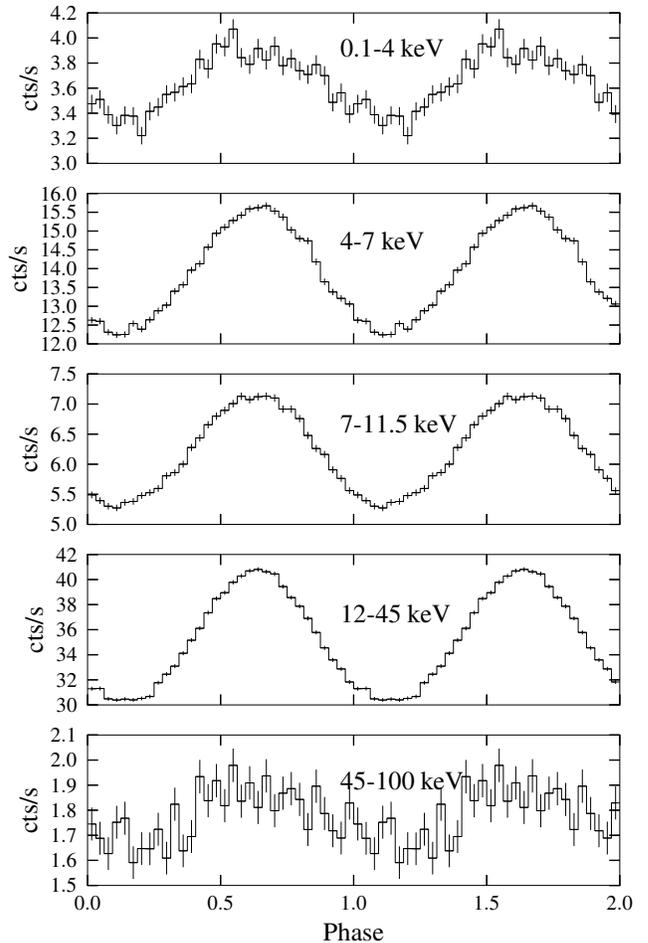}
	\caption{Pulse profiles of the pulsar \gro in 1997-03-21 for LECS (0.1 -- 4     keV), MECS (4 -- 7 -- 11.5 keV) and PDS (12 -- 45 -- 100 keV). The spin period is $P = 0.467044(1)$\,s.}
	\label{fig:970321pp}
\end{figure}

To study coherent pulsations, we corrected the photon arrival time
for Doppler delays due to the orbital motion of the spacecraft and the pulsar
(using the orbital ephemeris of \citealt{Kouveliotou:1996p6636}). Using the
phase-connection technique [see for instance]\citep{Doroshenko:2010p902} we measure 
the spin period to be $P = 0.467044(1)$\,s, $P = 0.467050(1)$\,s and $P = 0.46705(1)$ for the March and the two April
1997 observations (all uncertainties are for 90\% confidence level unless
stated otherwise). Folding the light curves reveals sinusoidal, single-peaked
pulse profiles (Fig. \ref{fig:970321pp}). We find that the phase of the pulsed
profiles during the bursts is shifted by $\sim5$\% relative to the persistent
state, which is consistent with earlier reports by
\cite{Kouveliotou:1996p6636,Strickman:1996p6385}. The morphology of the pulse
profiles appears to be consistent between the two observations and shows no
apparent dependence on luminosity.

\begin{figure}
	\includegraphics[width=0.5\textwidth]{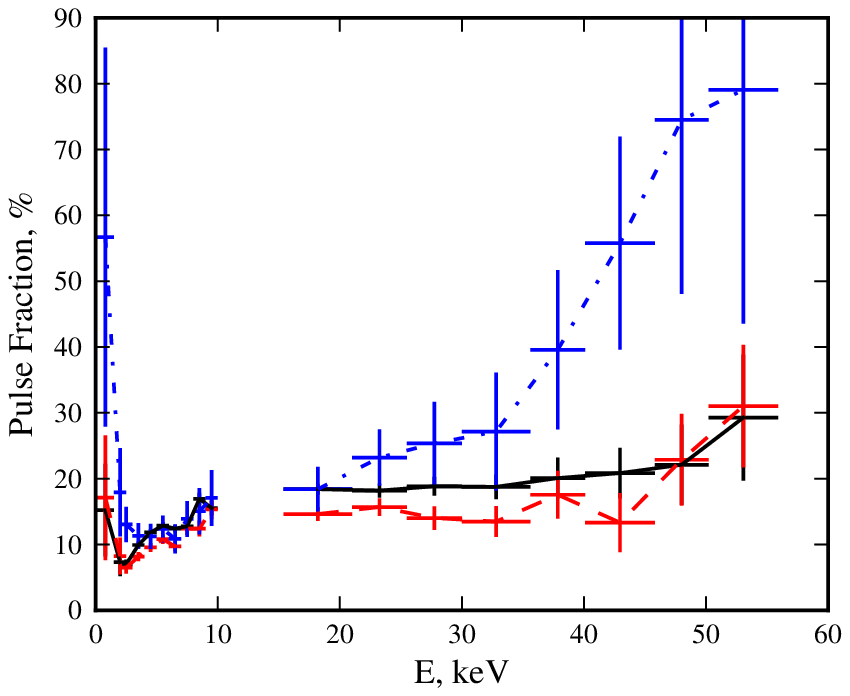}
	\caption{Pulse fraction of the pulsar \gro for 1997-03-21 - solid black         line, 1997-03-27 - dashed red line, 1997-04-12 - dashed dotted blue line.}
	\label{fig:pfrac}
\end{figure}

The fraction of pulsed emission defined as the ratio $(F_{\rm max} - F_{\rm min})/(F_{\rm max} + F_{\rm min})$, where $F_{\rm max}$ and $F_{\rm min}$ are the maximum and minimum source flux does change with energy. To
estimate it we assumed standard backgrounds for all instruments and also took
the contamination from the nearby sources to the PDS data (see
Section~\ref{spe}). We found that the pulsed fraction increases between 2 and
10 keV from about 5\% to 20\% (with a notable exception of a region close to
the fluorescent iron line, see also \citealt{Dai.etal:15}), and then remains
fairly constant between $\sim 10\%$ and $\sim 30\%$ for the first two
observations and between 20 -- 80\% for the third one. From 0.1 to 2 keV, the
pulse fraction decreases from 20 -- 30\% to 5\% although
our findings are at the lowest energies and for the third observation are hampered
by low statistics (Fig.~\ref{fig:pfrac}).

\subsection{Spectral analysis.}
\label{spe} \gro is located in the crowded Galactic Center region and a number
of sources fall into the field of view of the telescope. This might potentially
affect the spectral analysis, particularly for non-imaging instruments. Even
the LECS and MECS observations could be affected as the deep \emph{Chandra} observation of
the field reveals $\sim40$ sources within the standard 4\arcmin\, extraction
region. However, \gro is by far the brightest among these sources even in
quiescence, so for the imaging instruments MECS and LECS source confusion is
not an issue. We followed, therefore, standard analysis procedure.

\begin{figure}
	\includegraphics[width=0.5\textwidth]{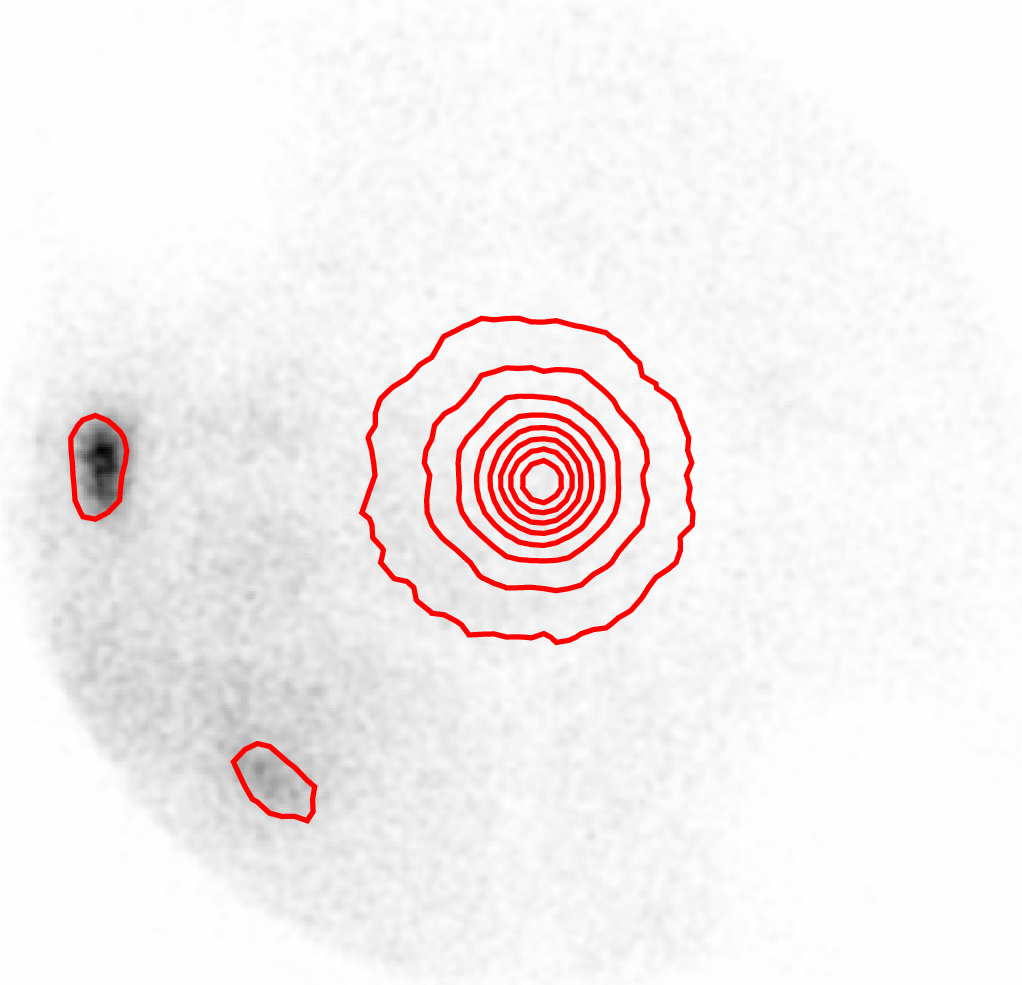}
	\caption{MECS image centered at \gro in quiescence on 1998-04-02 overlaid 
	with contours from an observation in bright state (taken on 1997-03-21). 
	Two additional sources detected in the MECS field of view contaminate also      the PDS spectrum.}
	\label{fig:mecs_ima}
\end{figure}

The situation for the non-imaging instruments is more complicated. The MECS image
reveals at least two sources besides \gro within the field of view of
collimating instruments. These are detected irrespective of the brightness of
the pulsar (see Fig. \ref{fig:mecs_ima}) and are plausibly identified as
1E~1743.1-2843 and Sgr~A*. To account for the in-orbit instrumental background which is rapidly changing and the for the contribution of the sources contaminating
the PDS data, we extracted the background-subtracted spectra for all observations
(including those in quiescence) using the standard pipeline. The standard pipeline allows to
account for instrumental in-orbit background. Then we subtracted quiescent spectra (obtained during the 1998 observations) from the source
spectra obtained for the first three observations. Apparently, an important
assumption here is that the flux of the contaminating sources remains constant.
To verify that we compared the soft X-ray fluxes measured by the MECS which
indeed remained constant.

For spectral fitting, we used the XSPEC package \citep[version 11.3.2,][]
{Arnaud:1996p6889}. We found that the observed broadband continuum spectrum can
be described with several phenomenological models generally used for the
spectra of accreting X-ray pulsars:

1) a power law plus a high-energy cutoff model, where the transition is smoothed with a Gaussian
$$
{\rm \mathit{PHC}}~\sim E^{-\Gamma}
\begin{cases} 
	1 & (E \le E_{\rm cut})\\ e^{-(E-E_{\rm cut})/E_{\rm fold}} & (E > E_{\rm cut})
\end{cases};
$$

2) a power law with Fermi-Dirac cutoff:
$$
 {\rm FDCO} \sim E^{-\Gamma}\, (1+e^{(E-E_{\rm cut})/E_{\rm fold}})^{-1};
 $$
 
3) the Negative and Positive power law EXponential model \citep{Mihara:1995p12, Makishima:1999p22}: 
$$
{\rm NPEX} = (A_1E^{-\alpha_1} + A_2E^{+\alpha_2})\, e^{-E/E_{\rm fold}};
$$

4) the \emph{CompTT} model by \cite{Titarchuk:1994p6894} describing the comptonization of soft photons with temperature $kT_{\rm s}$ in a hot
electron plasma with $kT_{\rm e}$ and a hot electron optical depth $\tau_{\rm e}$.

\begin{figure}
	\includegraphics[width=0.5\textwidth]{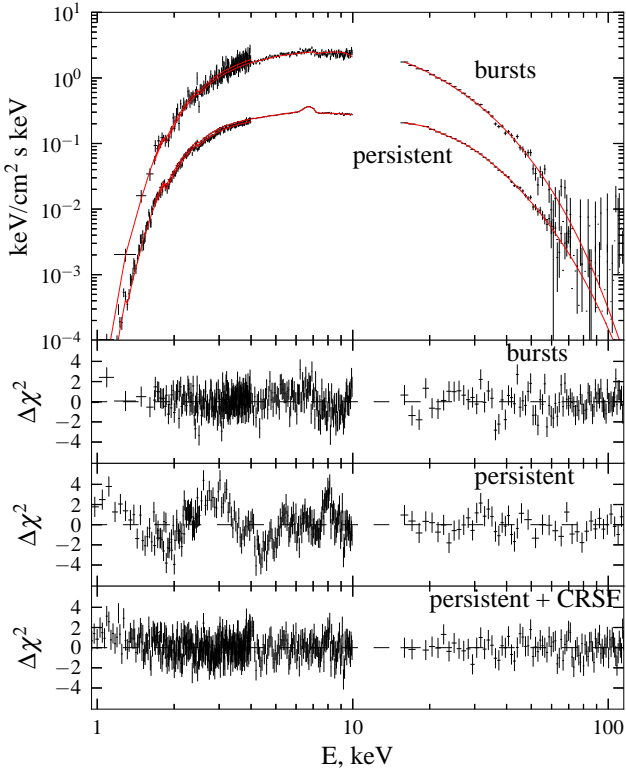}
	\caption{The best-fit unfolded average persistent spectrum, modelled with the \emph{PHC} continuum is shown in the top panel.  
	We also present the best fit residuals without and with the inclusion of a cyclotron line at $\sim 4.5$ keV. The best fit spectrum of the bursts modelled with \emph{PHC} and a black body 
	continuum is also presented together with the residuals for the observation 1997-03-21.}
	\label{fig:spe970321}
\end{figure}

\begin{figure}
	\includegraphics[width=0.5\textwidth]{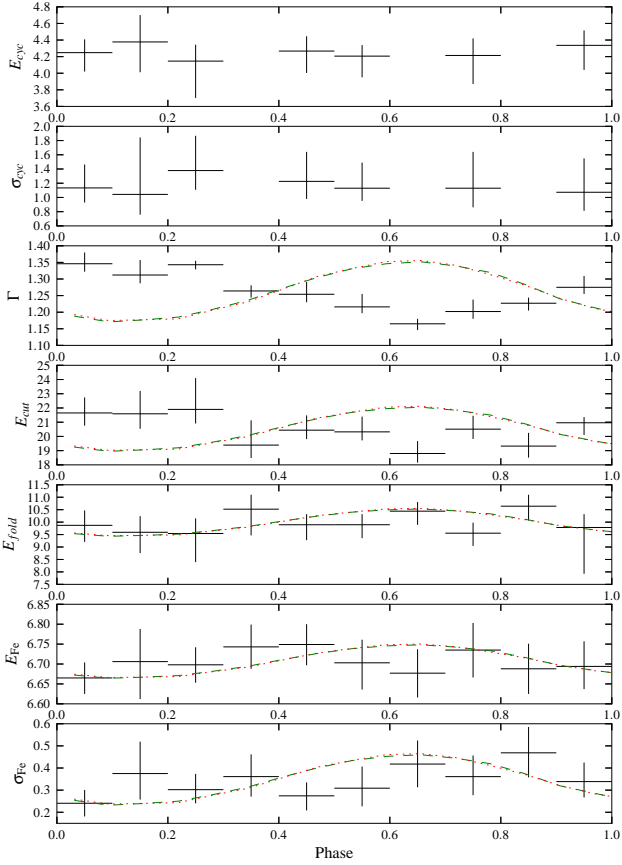}
	\caption{Spectral parameters of the pulsar \gro in 1997-03-21 with 
	the \emph{PHC} continuum model as function of pulse phase. Phase: 0-0.2-0.4-0.6-0.8-1. 
	Dashed lines are the MECS and PDS pulse profiles, they are the same.}
	\label{fig:phres21}
\end{figure}

In all cases the inclusion of the fluorescent iron K$_\alpha$ line at
$\sim6.7$\,keV (modelled with a simple Gaussian profile) was also required. All
continuum models give a reasonably good $\chi^2_{\rm red}$ (from 0.94 for the
\emph{PHC} model to 1.25 for \emph{CompTT} model, see Table \ref{tab:spe_p_all}). To
demonstrate the evolution of the parameters with the luminosity we used \emph{PHC}
based on the formal statistical quality and stability of the fit (Table
\ref{tab:spe_p}). Regardless of the assumed continuum model, some residuals
below 1 keV, and around 4--5 keV (Fig.~\ref{fig:spe970321}) were clearly observed. 
The absorption-like feature at $\sim 4$\,keV was modelled with a multiplicative
line with either a gaussian or lorentzian profile. This significantly
improved the fit (see  residuals of Fig. \ref{fig:spe970321}). The parameters of the line for the gaussian profile
are presented in Table~\ref{tab:spe_p_all}. For the lorentzian profile (CYCLABS in
XSPEC) we get comparable values, i.e. E$_{cyc} = 4.0 \pm 0.2$~keV, $\sigma =
1.3 \pm 0.2$~keV. The chance probability for fit improvement based on the
f-test is marginal $\le10^{-30}$ (see, however, \citealt{Protassov02}). The
feature also appeared in the phase resolved spectra (see Fig.
\ref{fig:phres21}). Our findings nicely agrees with the absorption-like feature recently reported by
\cite{Dai.etal:15} based on \emph{XMM-Newton} and \emph{INTEGRAL} observations
of the 2014 outburst. We note that \cite{Younes.etal:15} detected an absorption feature at $10$\,keV
which is not observed in our data, although the lack of HPGSPC data
makes it difficult to draw a definitive conclusion.

\begin{figure}
	\includegraphics[width=0.5\textwidth]{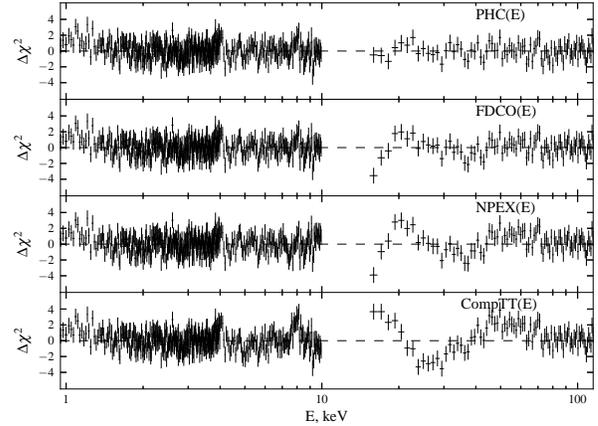}
	\caption{The best-fit residuals for spectra derived from the observation 
	1997-03-21 modeled as described in the main text. The best residuals obtained with the 
	power law plus high-energy cutoff model, \emph{PHC}(E), at the top panel.}
	\label{fig:speresid}
\end{figure}

\begin{table}
	\caption{Spectral parameters of the persistent spectrum of 
	\gro observed by \emph{BeppoSAX} in 1997-03-21 for various spectral models.
	All energies and line widths are given in keV. }
\begin{center}
\begin{tabular}{p{0.7cm}lllll}
\hline\hline
Par. & \emph{PHC} & \emph{PHC} & \emph{FDCO} & \emph{NPEX} & \emph{CompTT}\\
& & no CRSF & && \\
\hline
$N_{\rm H}^{\rm a}$ & 5.7(2)& 5.23(2)&5.55(8)& 5.16(6) &5.84(3)\\
$\Gamma$& 1.26(7)& 1.147(3)&1.09(4) & &\\
$E_{\rm cut}$& 18.4(1)& 18.1(1)&15(1) & &\\
$E_{\rm fold}$& 11.7(4)& 11.76(7)&10.2(2)& 6.1(1) &\\
$\alpha_1$ & & & &0.48(4) & \\
$\alpha_2$ & & & & -2.0 & \\
kT$_{\rm s}$ & & & & & 0.18(3) \\
kT$_{\rm e}$ & & & & & 5.82(3) \\
$\tau_{\rm e}$ & & & & & 13.1(1) \\
$E_{\rm Fe}$& 6.69(2)& 6.71(2)&6.70(2)& 6.68(2) & 6.72(2)\\
$\sigma_{\rm Fe}$& 0.28(3)& 0.38(2)&0.26(3)& 0.23(3) & 0.34(3) \\
$N_{\rm Fe}^{\rm b}$& 7.2(6) & 10.2(4)&5.8(5) & 5.8(6) &8.8(5) \\
$E_{\rm cyc}$& 4.3(2) & &4.47(9)& 4.55(5) & 4.2(3)\\
$\sigma_{\rm cyc}$ & 1.2(3)& &1.1(2)& 1.2(1) & 0.7(5)\\
\hline
$\chi^2_{\rm red}$/dof &0.94 / 549 & 1.28 / 552 &0.95 / 551 & 1.04 / 551 & 1.25 / 553\\
\hline
\end{tabular}
\end{center}
$^a${in units $10^{22}$ atoms  cm$^{-2}$}\\
$^b${K$_\alpha$ line normalization in units  10$^{-3}$ph cm$^{-2}$ s$^{-1}$}
\label{tab:spe_p_all}
\end{table}

\begin{table}
	\caption{Spectral parameters of the persistent spectrum low mass X-ray pulsar 
	\gro observed by \emph{BeppoSAX} modeling by the \emph{PHC} with smoothing gaussian continuum model.
	All the energies and the line widths are given in keV. }
\begin{center}
\begin{tabular}{llll}
\hline\hline
Parameter &1997-03-21& 1997-03-27& 1997-04-12\\
\hline
$N_{\rm H}^{\rm a}$ & 5.7(2)& 5.7(2)& 5.4(1)\\
$\Gamma$& 1.26(7)& 1.29(5)& 1.24(3)\\
$E_{\rm cut}$& 18(1)& 19.3(8)& 24(4)\\
$E_{\rm fold}$& 11.7(4)& 13.3(5)& 12(3)\\
$E_{\rm Fe}$& 6.69(2)& 6.70(3)&6.68(5)\\
$\sigma_{\rm Fe}$& 0.28(3)& 0.27(4)& 0.24(7)\\
$N_{\rm Fe}^{\rm b}$& 7.2(7) & 4.9(6) & 1.2(2) \\
$E_{\rm cyc}$& 4.3(2) & 4.3(1)& \\
$\sigma_{\rm cyc}$ & 1.2(3)& 1.1(1)& \\
$F_{\rm ab}^{\rm c}$ & 10.1 & 6.5 & 1.2 \\
$F_{\rm unab}^{\rm c}$  & 13.0 & 8.3 & 1.6 \\
\hline
$\chi^2_{\rm red}$/dof &0.94 / 549 & 1.04 / 531 & 1.06 / 487\\
\hline
\end{tabular}
\end{center}

$^a${in units $10^{22}$ atoms  cm$^{-2}$}\\
$^b${K$_\alpha$ line normalization in units  10$^{-3}$ph cm$^{-2}$ s$^{-1}$}\\
$^c${Absorbed and unabsorbed fluxes in units $10^{-9}$ erg  cm$^{-2}$\, s$^{-1}$ and in 0.1 -- 120 keV range.} 
\label{tab:spe_p}
\end{table}

We have also performed a pulse phase resolved spectral analysis to study the
variation of spectral parameters with the angle of view to the neutron star.
Only data from the observation on March 21, 1997 with highest luminosity and
best counting statistics was considered. To describe the phase resolved
spectra, we used the same model as for the phase averaged spectrum, i. e. a
power law with high-energy cut-off (see Fig. \ref{fig:phres21}). We divided the
data in 10 equally spaced phase bins. We fixed the absorption component at the
average value $N_{\rm H} = 5.7_{-0.2}^{+0.2} \times 10^{22}$atoms cm$^{-2}$.
The emission iron line at $\sim 6.7$ keV and the absorption-like feature at
$\sim 4.5$ keV were also included in the phase resolved analysis. We can see
the clear anti-correlation of the photon index with phase flux. The parameters
of the absorption-like and iron lines (line centroid and width) did not exhibit
any significant phase variation. The cut-off and folding energies appeared to
be anti-correlated with each other.

\subsubsection{Analysis of the burst spectra}
To investigate the burst spectra we aligned and stacked all observed bursts
using the rising edge as a reference to improve statistics. We also subtracted
the contribution of the persistent emission. We found that the shape of the
combined spectrum of the bursts departs significantly from that of the
non-bursting spectrum (Fig.\ref{fig:spe970321}).

We first described the burst spectrum with the \emph{PHC} and
\emph{CompTT} models. Both models give formally acceptable results
(see Table\,\ref{tab:spe_b}). The spectral parameters of the \emph{PHC} model
are similar to the ones reported recently by \cite{Younes.etal:15} for the burst spectra ($\Gamma = 0.2 \pm 0.1$ and $E_{\rm fold} = 7.6\pm
0.5$\,keV). Such a hard power law together with a soft cut-off are rather peculiar and not typical of X-ray pulsars. This might suggest that 
the model describes a bump-like feature at soft energies. 
In addition, the best-fit results for the bursts spectrum using the \emph{CompTT} model reveals a significant change, with respect to the persistent spectrum, of both the seed
photon temperature and the absorption column (seeTable~\ref{tab:spe_b}), which is difficult to understand.  On the other hand,
the apparent change of the seed photon temperature might suggest an
additional soft component in the burst spectrum. Indeed, including a
blackbody component in either model allows to achieve comparable (or, in fact,
slightly better) fit statistics while the other parameters of the continuum remain
close to ones measured for the persistent spectrum. In other words, we find
that the burst spectrum significantly differs from the persistent one
at soft energies, and this change is well accounted for by adding
a soft blackbody-like component to the unchanged continuum of the persistent spectrum. The blackbody component has a temperature about $2.1-2.2$ keV and a
size close to a neutron star radius (for an assumed distance of $8$\,kpc). A Fe-K line
is also marginally significant in the burst spectra of the observation on
1997-03-27. We have summarised the fit results for the 1997-03-21 data in
Table~\ref{tab:spe_b}.

\begin{table}
	\caption{Best-fit parameters of the bursts spectrum \gro in 1997-03-21. 
	All the energies and the line widths are given in keV.
	}
\begin{center}
\begin{tabular}{lllll}
\hline\hline
Parameter & \emph{PHC} &\emph{bb+PHC}& \emph{CompTT} & \emph{bb+CompTT}\\
\hline
N$_{\rm H}^a$& 5.0(3)  &5.1(7)& 3.4(2)& 5.2\\
kT$_{\rm bb}$& & 2.1(3)&& 2.2(1)\\
R$_{\rm bb}^b$ &  &$\sim 7.5$& & $\sim 7.6$\\
$\Gamma$& 0.26(8)& 0.8(4)& &\\
E$_{\rm cut}$& 4(1) & 18(5)&  &\\
E$_{\rm fold}$& 7.6(2) & 9(1) &&\\
kT$_{\rm s}$& & &1.51(8)&0.2\\
kT$_{\rm e}$& & &5.4(2)& 5.2(1)\\
$\tau_{\rm e}$&&& 4.6(5)& 18(1)\\
\hline
$\chi^2_{\rm red}$/dof &0.96 / 408 & 0.95 / 406 & 1.03 / 409 & 1.02 / 409\\
\hline
\end{tabular}
\end{center}

$^a${in units $10^{22}$ atoms  cm$^{-2}$}, 
$^b${in km.} 
	The absorbed and unabsorbed fluxes in 0.1 -- 120 keV energy range are 
	F$_{\rm ab} = 6.09 \times 10^{-8}$ erg cm$^{-2}$ s$^{-1}$, F$_{\rm unab} = 6.83 \times 10^{-8}$ erg cm$^{-2}$ s$^{-1}$.
\label{tab:spe_b}
\end{table}

\begin{figure}

	\includegraphics[width=0.5\textwidth]{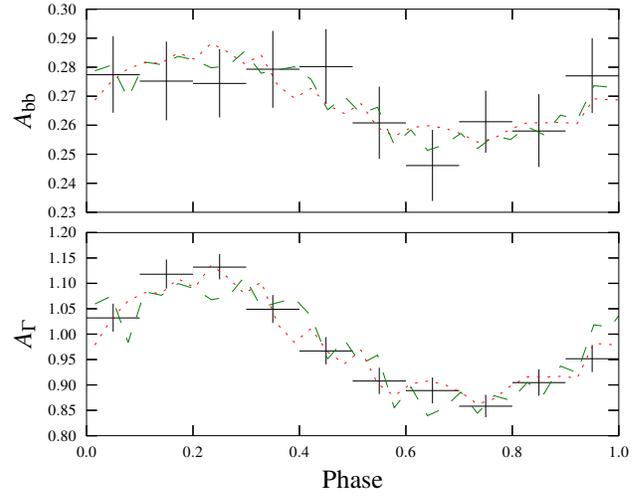}
	\caption{Changing normalizations of the soft (top) and hard (bottom) part of the bursts spectrum 
	with phase for the \gro in 1997-09-21. $\chi^2_{\rm red}$ lies in region from 
	0.97 to 1.45.}
	\label{fig:PheResBursts}
\end{figure}

To investigate the pulse phase dependence of the black body and power law
components we carried out the phase resolved analysis of the burst spectrum.
Unfortunately, the statistics was insufficient to constrain all model parameters,
so we kept most of the parameters fixed to the best-fit phase average values,
and only allowed the normalizations of the continuum components to vary. The
results are shown in Fig.~\ref{fig:PheResBursts}. The pulse fraction for the
soft and hard components are $6.5\% \pm 3.4\%$ and $13.8\% \pm 1.9\%$
respectively, i.e. the hard component varies with pulse-phase stronger than the
soft one.

\begin{figure}
	\includegraphics[width=0.5\textwidth]{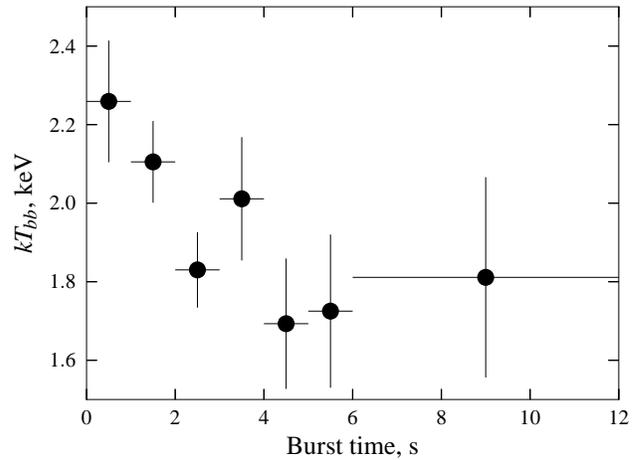}
	\includegraphics[width=0.5\textwidth]{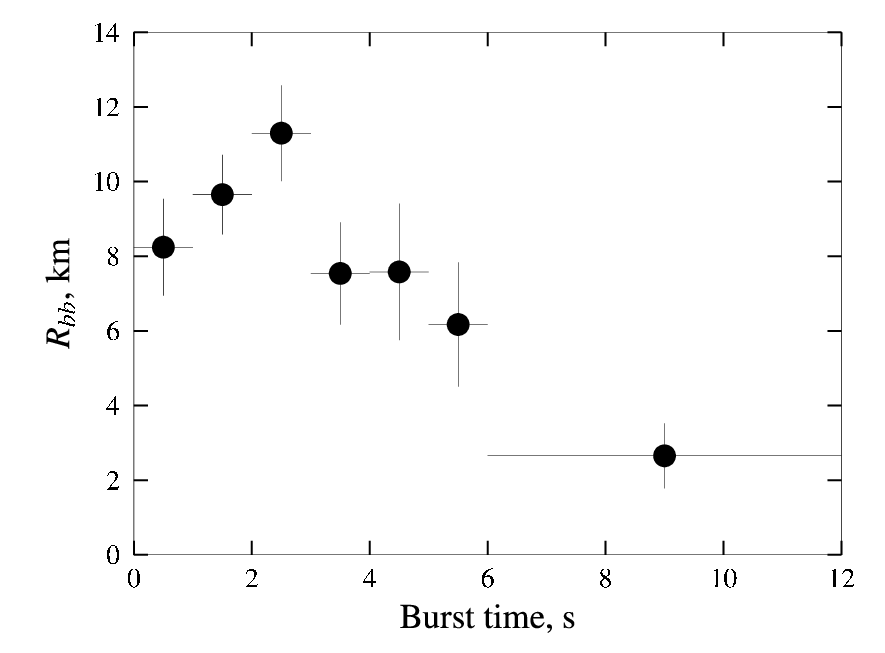}
	\caption{Cooling of the soft component, during 
	the bursts. The probability that the observed temperatures are due to a statistical
    fluctuation is $\le1.7\times10^{-8}$ (from Kolmogorov-Smirnov test), so the trend is significant.}
	\label{fig:bb_cooling}
\end{figure}

To explore the time evolution of both components along the burst, we
performed also time-resolved spectral analysis using the stacked data of all
bursts. It is interesting to note, that both the temperature and normalization of
the blackbody component (and thus the size of the emitting region) change along
the burst as shown in Figure~\ref{fig:bb_cooling}. 

\section{Discussion} The non-bursting spectrum of \gro has been measured for the
first time in a broad energy range 0.7 to 120 keV with \emph{BeppoSAX} during the source
outburst in 1996-1997. The persistent spectrum of the source is similar to the one of accreting
pulsars and can be described by phenomenological models typically used for pulsars, e.g. by an absorbed power law with cut-off at around $\sim 18$\,keV. 
In addition, a line-like absorption feature at $E\sim4.5$\,keV is observed and required regardless of the used continuum model. 
We incidentally observe that there are no systematic effects known for the MECS and LECS, in the energy range where residuals
are observed. In addition, this absorption-like feature is not observed in the
spectrum of the bursts, which has similar shape and comparable statistical
quality. Therefore, the feature is unlikely to be an artefact of the continuum modelling.

Cyclotron resonance scattering features (CRSFs), interpreted as a result of
resonant scattering of continuum photons off plasma electrons in strong
magnetic fields close to the electron gyro frequency, are often observed in the spectra
of accreting pulsars. CRSFs appears as a flux suppression of the continuum at cyclotron energies
related to the magnetic field strength in the line forming region:
$$ 
E_{\rm cyc}\simeq11.57\,B_{12}\,(1+{\rm z})^{-1}\,\,{\rm keV},
$$
where $B_{12}$ is the magnetic field strength in the units of $10^{12}$~G, z is
the gravitational redshift (see \cite{crsf_review}, \cite{Doroshenko:2010p902},
\cite{Santangelo:1999p314} for a recent review). By interpreting an absorption-like feature observed at $E\sim 4.5$~keV in the spectrum of \gro as a CRSF,
a magnetic field of $B\sim3.7 \times 10^{11}(1+\rm z)$\,G is obtained for the neutron star of the binary system.  
This is in strikingly good agreement with earlier estimates of the magnetic field of this pulsar, which fall in range
2-7$\times 10^{11}$\,G \citep{Finger:1996p6585,Rappaport:1997p6322,Cui:1997p6334} and \citep{Younes.etal:15,Dai.etal:15}. 

In particular, \cite{Dai.etal:15} have recently reported the detection of such an absorption
feature at $\sim4.7$\,keV based on \emph{XMM-Newton} and \emph{INTEGRAL} data,
which they also interpret as a cyclotron resonance scattering feature (CRSF). Our result is in excellent agreement with these findings. 

Besides the lack absence of absorption feature, the burst spectrum was found to differ
significantly from the non-bursting spectrum, particularly at soft energies.
The difference is most easily accounted for by inclusion of a black-body
component on top of the power law-like persistent spectrum of accreting pulsars. The flux in 0.7 --
120 keV energy range measured for the power law and the blackbody components is
F$_{\rm bb} \approx 1.6 \times 10^{-8}$ erg cm$^{-2}$ s$^{-1}$ and F$_{\rm
power} \approx 4 \times 10^{-8}$ erg cm$^{-2}$ s$^{-1}$, respectively. This indicates that 
$\sim30$\% of the energy released in bursts arises from the soft thermal
component not present in the persistent emission.

The nature of the soft component is unclear and several hypotheses may be
considered. First of all, spectra of accreting pulsars are known to change with
luminosity and the luminosity in \gro increases by an order of magnitude during
the bursts, so some intrinsic variation of the spectrum might be anticipated.
For instance, \cite{Reig13} reported evident spectral softening for several
pulsars above the so-called critical luminosity $\sim10^{37}$\,erg\,s$^{-1}$
\citep{BS76,Mushtukov15}, and similar softening might occur in \gro, which at
$L_x\sim10^{38}$\,erg\,s$^{-1}$ is likely also above this limit.

Emission from the accretion column expected to form at this stage
can also irradiate the accretion disk or the surface of neutron star
\citep{LS88, Poutanen.etal:13}, and the thermalization of intercepted emission
could be responsible for the soft component as well. The effective temperature
of the irradiated surface $T_{\rm eff} \approx (L_{\rm x}/4\pi r^2/\sigma_{\rm
SB})^{1/4}\sim 2$\,keV is, in fact, comparable with the observed. On the other
hand, strong gravitational separation of chemical elements in the neutron star
atmospheres \citep{Hameury.etal:83} implies that the upper atmospheric layers
consist almost exclusively of fully ionized hydrogen plasma which reflects most
of the incident flux due to electron scattering. Therefore, the fraction of
thermalized emission is likely to be less than 10\% \citep[see a detail
argumentation in ][]{Poutanen.etal:13}, in contrast with the $\sim30\%$
observed. It is also hard to explain the observed pulsation of the soft
component if it originates in the inner regions of accretion disk, although the
even softer component with $kT\sim0.5$\,keV reported recently by
\cite{Younes.etal:15, Dai.etal:15} might indeed be related to irradiation of
the accretion disk by the pulsar.

Finally, taking into consideration the observed cooling of the thermal
component along the burst (see Fig.\,\ref{fig:bb_cooling}) very similar to that
observed in classical bursters \citep{LvPT:93} thermonuclear flashes on the NS
surface could be also responsible for the observed soft emission. In fact,
several authors have already considered the possibility that some of the bursts
in \gro might be Type-I bursts. 
Based on the \emph{BATSE} observations, \citet{Lewin:1996p469} concluded that this
is unlikely because the amount of
matter accreted between the bursts is insufficient to explain the observed
burst fluence if bursts are powered by thermonuclear burning. Indeed, the
inter-burst to burst fluence ratio $\alpha\sim4$ deduced from \emph{BATSE}
observations was found to be much smaller than $\alpha\ge40$ typical of
thermonuclear hydrogen burning \citep{LvPT:93}.

On the other hand, \cite{Jahoda97}, based on the \emph{RXTE} observations close
to the peak of the outburst, found significanly higher $\alpha\sim34$. Moreover,
we found that the flux of the thermal component constitutes only about 30\% of
the total flux and, and therefore of the burst fluence. If this is taken into
account, even at lower luminosities such as during the \emph{BeppoSAX}
observations, when we estimate $\alpha\sim 5-15$ for the bolometric flux, the
same ratio calculated for the thermal component alone is factor of three
higher, i.e. still comparable with that observed in classical bursters.
Therefore the argument by \cite{Lewin:1996p469} clearly does not always hold.

The strongest argument against the thermonuclear origin of the soft component
in burst spectrum of \gro is that it is an accreting pulsar. Indeed, the
temperature and pressure of matter funneled by the magnetic field to the polar
areas are usually sufficient for stable thermonuclear burning, so no unburned
matter accumulates at the surface and thus no Type-I bursts are observed from
accreting pulsars. In fact, \cite{BB:97} did consider the possibility of unstable
thermonuclear burning in \gro, and concluded that under the most favourable
assumptions it should not be possible. In particular, \cite{BB:97} argue that
to escape the accretion flow and spread over the NS surface (where it can be
accumulated and subsequently ignite as a thermonuclear flash), the accreted
plasma must overcome the magnetic pressure and turn the magnetic field lines
parallel to the neutron star surface. This implies a pressure of $P \approx
10^{24}$\,erg\,cm$^{-3}$ while the hydrogen/helium mixture burns in the stable
regime at $P \approx 10^{22}$\,erg\,cm$^{-3}$. 

In other words, accreting matter remains confined at the polar caps where it
burns steadily as expected. However, in our opinion, this key assumption of
\cite{BB:97} is probably too conservative, and, in fact, corresponds to the
case when the plasma is not confined to polar areas at all. This would
basically imply that \gro is an ordinary burster, which clearly it is
not. On the other hand, the magnetic pressure at the poles of the neutron star
can be estimated based on the observed CRSF energy and turns out to be
$B^2/8\pi \approx 10^{22}$\,erg\,cm$^{-3}$ comparable with the pressure
required for stable thermonuclear burning. Therefore, the magnetic field is
unlikely to confine plasma at significantly higher pressures and part of
accreted matter probably leaves the polar areas before burning, thereby
creating conditions for thermonuclear flashes. We note also that, if realised,
the onset of a such flash could in principle disrupt the inner parts of the
accretion disk triggering the instabilities responsible for the enhanced
accretion rate and most of the observed burst emission.

In conclusion, we fully agree that there is no doubt that the mass accretion
rate does increase during the bursts \gro and is responsible for the bulk of
the observed emission. However, some fraction of the flux in principle could
still be due to unstable thermonuclear burning at the NS surface.

\section{Summary and conclusions}
We presented the results of the analysis of three \emph{BeppoSAX} observations, with
a total exposure time 270\,ks, carried out in the declining phase of the 1997
outburst of the unique bursting pulsar \gro.

Pulsations with a period of 0.4670\,s with a stable pulse profile were detected in
all observations. The pulsed fraction was found to vary with energy, reaching
a minimum of 10$-$20\% in the energy range 5$-$40\,keV and increasing for higher
and lower energies, especially at lower luminosities. Several tens of bursts
with typical durations of about 10\,s were detected as well. Depletion in the
light curve is observed after at least some of the bursts. The source
luminosity typically increases by factor of ten during the bursts,
although a number of dimmer bursts are also observed.
    
The non-bursting broad band X-ray spectrum was found to be well described by
several phenomenological models typically used for accreting pulsars. An iron
line at $\sim$\,6.7\,keV and an absorption feature at $\sim$\,4.5\,keV were also
required to fit the data. We interpret the absorption feature as a cyclotron
line which implies a magnetic field of the neutron star of $B\sim 3.7\times10^{11}$\,G. This value
is in good agreement with earlier predictions and with the recent observations of the 2014
outburst with \emph{XMM-Newton} and \emph{INTEGRAL}. \cite{Dai.etal:15} 
have detected the same feature at $E\,\sim4.7$\,keV in agreement, within uncertainties,
with our measurement.

The average burst spectrum could be represented as a combination of the harder
non-bursting spectrum and an additional soft thermal component with temperature of
about 2\,keV. The burst spectrum requires that neither the iron 6.7\,keV line nor
the absorption feature at 4.5\,keV. Both components are pulsed although the
amplitude is smaller for the soft component, $\sim$\,6.5\% vs $\sim$\,13.8\%.
    
We discussed a possible nature of the thermal component and speculate that it
could be caused by thermonuclear flashes which possibly trigger the accretion
rate enhancement responsible for the bursts observed in the source. This
hypothesis is based on several similarities between typical Type I bursts and
the thermal component, namely, the burst duration of 10\,s, the inter-burst to burst
fluence ratio and cooling as a burst progresses. There are some observational
and theoretical arguments that disfavor this hypothesis. These cannot be
verified with the existing data, however, additional data from current or
future missions like LOFT \citep{loft} might provide decisive insights on this
puzzling source.

\section*{Acknowledgements}
This work is partially supported by the
Bundesministerium für Wirtschaft und Technologie through the Deutsches
Zentrum für Luft und Raumfahrt (DLR, Grants FKZ 50~OG~1001,
50~OR~0702, 50~QR~1008), the German Research Foundation (DFG) grant
WE 1312/48-1.
\bibstyle{mn2e}
\bibliography{bib_clean1}
\end{document}